\documentclass[aps,pre,twocolumn,amsmath,amssymb,floatfix,footinbib,superscriptaddress,letterpaper]{revtex4-1}

\usepackage{amsmath,amssymb,mathtools}
\usepackage{stmaryrd}
\usepackage{times}
\usepackage[euler]{textgreek}
\usepackage[bbgreekl]{mathbbol}
\usepackage{graphicx}
\usepackage{bm,color}
\usepackage{natbib,hyperref}
\usepackage[capitalize]{cleveref}

\hypersetup{colorlinks=true,linkcolor=blue,citecolor=blue,urlcolor=blue}

\begin{document}
\title{Mass sensing by detecting the quadrature of coupled light field}
\author{Qing Lin}
\affiliation{Fujian Key Laboratory of Light Propagation and Transformation, College of Information Science and Engineering, Huaqiao University,
Xiamen 361021, China}
\author{Bing He}
\affiliation{Department of Physics, University of Arkansas, Fayetteville, AR 72701, USA}
\author{Min Xiao}
\affiliation{Department of Physics, University of Arkansas, Fayetteville, AR 72701, USA}
\affiliation{National Laboratory of Solid State Microstructures and School of Physics, Nanjing University, Nanjing 210093, China}

\begin{abstract}
Ultrasensitive detections have been proposed as an application of optomechanical systems. Here we develop an approach 
to mass sensing by comparing the detected quadratures of light field coupled to a mechanical resonator, whose slight change of the mass should be precisely measured. The change in the mass of the mechanical resonator will cause the detectable difference in the evolved quadrature of the light field, to which the mechanical oscillator is coupled. It is shown that the ultra-small change $\Delta m$ from a mass $m$ can be detected up to the ratio $\Delta m/m\sim 10^{-8}-10^{-7}$ by choosing the feasible system parameters.
\end{abstract}

\maketitle
\section{Introduction}

Highly precise detection is an important ingredient in modern technologies. One category is the precise measurement of the masses of nanoparticles or biomedical molecules, the applications of which cover early-stage disease diagnosis, environmental monitoring, emergency response, and homeland security \cite{nbt1, nbt2, nbt3}. Due to the possibility of realizing an ultra-high quality factor $Q_m$ of mechanical resonator, for example $Q_m=10^8$ in microtoroidal cavity \cite{hq}, the coupled systems of mechanical resonator with cavity field was regarded as a good candidate for various precise measurements \cite{chen}, including the detection of gravitational waves \cite{ligo1, ligo2,ligo3,ligo4,ligo5} which is currently under deep concern. As we will show below, the similar systems can be applied to detect a very small mass too.

Previously the detection of nanoparticles was mostly through their modification of the whispering gallery modes (WGMs) of optomechanically coupled systems \cite{WGMX}. When they are attached to cavity, the resonance frequency of WGMs will be shifted according to the size of the nanoparticles. By detecting the shift of WGMs, the ultra-sensitive size sensing can be realized \cite{WGM1,WGM2,WGM3}. In addition, the dispersion of nanoparticles will induce the mode splitting \cite{WGM4,WGM5,WGM6,WGM7,WGM8,WGM9,WGM10,WGM11,WGM12,WGM13,WGM14,WGM15,WGM16} 
or mode broadening \cite{WGM17,WGM18,WGM19}, and will also cause the linewidth change \cite{WGM20} which can be used for size sensing as well.

A further step is to find out the small masses of the nanoparticles with the similar systems. If a particle with the mass $\Delta m$ is attached to the mechanical resonator with the original mass $m$, its mechanical resonance frequency will be lowered by a small quantity 
$\Delta\omega_m$ according to the relation \cite{ms1}
\begin{align}
\Delta\omega_m=-\frac{\Delta m}{2m}\times\omega_m,
\end{align} 
where $\omega_m$ is its original resonance frequency.
This relation has been used to make ultra-sensitive mass sensors; 
see, e.g \cite{ms1,ms2,ms3,ms4,ms5,ms6,ms7,ms8,ms9,ms10,ms11,ms12,ms13,ms14,ms15,ms16,ms17,ms18,ms19,ms20}.

Here we present a feasible approach to the mass sensing by detecting the change in the quadrature of the coupled light field to the mechanical resonator, as the result of its variation of the resonance frequency by $\Delta\omega_m$ due to the extra particles. 
A slight change of the mechanical resonator's mass can lead to the detectable change in the coupled cavity field, or more exactly the amplitude or phase of cavity quadratures will change according to the modification $\Delta\omega_m$ to the system. We apply this mechanism to realize a mass sensing. Particularly we will show that the sensitivity can reach an ultra-high level $\Delta m/m=10^{-8}-10^{-7}$ by choosing the feasible system parameters. 

The rest of the paper is organized as follows. In Sec. \ref{sec:model}, we describe the model of the concerned system for the quantitative discussions in the following sections. The mechanism for the mass sensing by the comparison of cavity quadratures, which is detailed in Sec. \ref{sec:real}, is illustrated with the system's dynamical equations. Then we systematically investigate how the system parameters affect the detection precision in Sec. \ref{sec:increase}, in order to optimize the performance. The work is concluded in the final section.

\section{The model}
\label{sec:model}

\begin{figure}[b!]
\centering
\includegraphics[width=0.7\linewidth]{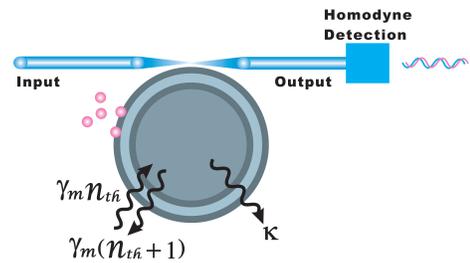}
\caption{Scheme for mass sensing with an optomechanically couped system. The nanoparticles are attached to the mechanical resonator, resulting in a change of the mechanical resonant frequency, which will affect the cavity quadratures through the interaction between the mechanical resonator and the cavity field. The change of the cavity quadratures can be detected by a homodyne detection on the output field. Here the mechanical resonator is initially in thermal equilibrium with its environment, which has the thermal occupation $n_{th}$.}
\label{fig1}
\end{figure}

The optomechanically coupled system used for detection is depicted in Fig. 1, where a mechanical resonator (the boundary of an expandable cavity) is driven by a continuous-wave (CW) laser field via the radiation pressure. The Hamiltonian, $H=H_S+H_{OM}+H_{SR}$, to describe the dynamical process due to the interaction between the cavity field and the mechanical resonator consists of three parts. 
The first one about the external drive and the system modes' oscillations takes the form ($\hbar=1$)
\begin{align}
H_S=\Delta \hat{a}^{\dag}\hat{a}+\omega_m \hat{b}^{\dag}\hat{b}+iE(\hat{a}^{\dag}-\hat{a})
\end{align}
in a rotation frame with respect to the external drive frequency $\omega_L$ \cite{re}, where $\Delta=\omega_c-\omega_L$ is the drive 
detuning, and $\omega_c$ ($\omega_m$) is the frequency of the cavity (mechanical) mode which oscillates under a CW laser with the 
constant amplitude $E$. The second one $H_{OM}=-g_{m}\hat{a}^{\dag}\hat{a}(\hat{b}+\hat{b}^{\dag})$ is the coupling of the cavity field with the mechanical oscillator due to the radiation pressure, where $g_{m}$ is the coupling constant at the single photon level. The final one
\begin{align}
H_{SR}(t)=&i\sqrt{2\kappa}\{\hat{a}^{\dag} \hat{\xi}_c(t)e^{i\omega_L t}-\hat{a}\hat{\xi}^{\dag}_c(t)e^{-i\omega_L t}\}\nonumber\\
&+i\sqrt{2\gamma_m}\{\hat{b}^{\dag} \hat{\xi}_m(t)-\hat{b}\hat{\xi}^{\dag}_m(t)\}
\label{st}
\end{align}
about the linear coupling between the cavity (mechanical) mode with the associated reservoir, which acts on the system via the stochastic Langevin noise operator $\hat{\xi}_c$ ($\hat{\xi}_m$), gives rise to the damping rate $\kappa$ ($\gamma_m$) of the system modes. The system is prepared in thermal equilibrium with the environment before the drive laser is turned on.

We adopt an approach of factorizing a system's evolution operator \cite{fqa1,opt,fqa2,other,fqa3,e1,fqa4,e2,qby1} to study the system. 
For the currently concerned evolution operator $U(t)=\mathcal{T}\exp\{-i\int_0^t d\tau \hat{H}(\tau)\}$, the reservoirs' action manifests by the stochastic Hamiltonian (\ref{st}) among the total Hamiltonian $H(t)$. 
First we take an interaction picture with respect to the system Hamiltonian $\hat{H}_S$, 
which is equivalent to a factorization of the evolution operator as
\begin{align}
U(t)=e^{-i H_S t}\times\mathcal{T}e^{-i\int_0^t d\tau \{H_{eff}(\tau)+H_N(\tau)\}},
\label{decomp}
\end{align}
where $H_{eff}(t)+H_N(t)=e^{iH_S t}\{H_{OM}+H_{SR}(t)\}e^{-iH_S t}$.
It results in the effective Hamiltonian
\begin{align}
H_{eff}(t)&=-g_m\left[E(t)\hat{a}^{\dag}+E^*(t)\hat{a}+|E(t)|^2\right]\nonumber\\
&\times(e^{-i\omega_mt}\hat{b}+e^{i\omega_mt}\hat{b}^{\dag}) \nonumber\\
&+i\sqrt{2\kappa}\{e^{i\omega_c t}\hat{A}^{\dag}(t) \hat{\xi}_c(t)-e^{-i\omega_c t}\hat{A}(t)\hat{\xi}^{\dag}_c(t)\} \nonumber\\
&+i\sqrt{2\gamma_m}(e^{i\omega_mt}\hat{b}^{\dag}\hat{\xi}_m(t)-e^{-i\omega_mt}\hat{b}\hat{\xi}^{\dag}_m(t)), 
\label{eff}
\end{align}
in addition to a nonlinear part $H_N=-g_m\hat{a}^{\dag}\hat{a}(e^{-i\omega_mt}\hat{b}+e^{i\omega_mt}\hat{b}^{\dag})$,
where
\begin{align}
\hat{A}(t)&\equiv e^{i \hat{H}_S t}\hat{a}e^{-i \hat{H}_S t}=e^{-i\Delta t}(\hat{a}+E(t)),\nonumber\\
E(t)&=\frac{iE}{\Delta}(1-e^{i\Delta t}).
\label{def}
\end{align}
In the currently concerned setups with $g_m/\omega_m\ll 1$, the effect of the nonlinear part $H_N$ can be well neglected 
as compared with the quadratic Hamiltonian $H_{eff}$ \cite{fqa4,qby1}.

The cavity field quadratures $\hat{X}_{c}(t)=(\hat{a}(t)+\hat{a}^\dagger(t))/\sqrt{2}$ and $\hat{P}_c(t)=i(\hat{a}(t)-\hat{a}^\dagger(t))/\sqrt{2}$ are thus determined by the system Hamiltonian $H_S$ and the effective Hamiltonian $H_{eff}$ according to the two combined actions in Eq. (\ref{decomp}). The former transform them 
to
\begin{align}
\hat{X}^{(1)}_c\equiv& e^{i \hat{H}_S t}\hat{X}^0_c e^{-i \hat{H}_S t}\nonumber\\
=&\cos(\Delta t)\hat{X}^0_{c}+\sin(\Delta t)\hat{P}^0_{c}\nonumber\\
&+(e^{-i\Delta t}E(t)+e^{i\Delta t}E^*(t))/\sqrt{2},\nonumber\\
\hat{P}^{(1)}_c\equiv& e^{i \hat{H}_S t}\hat{P}^0_c e^{-i \hat{H}_S t}\nonumber\\
=&-\sin(\Delta t)\hat{X}^0_{c}+\cos(\Delta t)\hat{P}^0_{c}\nonumber\\
&-i(e^{-i\Delta t}E(t)-e^{i\Delta t}E^*(t))/\sqrt{2}, \label{txp}
\end{align}
where $\hat{X}^0_c=(\hat{a}(0)+\hat{a}^\dagger(0))/\sqrt{2}$ and $\hat{P}^0_c=i(\hat{a}(0)-\hat{a}^\dagger(0))/\sqrt{2}$.
The latter will evolve the quadratures according to the linear dynamical equations:
\begin{align}
\dot{\hat{X}}^{(2)}_c=&-\kappa \hat{X}^{(2)}_c-2g_m\mathcal{I}_m(E(t))[\cos(\omega_m t)\hat{X}_m+\sin(\omega_m t)\hat{P}_m]\nonumber\\
&-\sqrt{2}\kappa\mathcal{R}_e(E(t))+\sqrt{\kappa}[e^{i\omega_c t}\hat{\xi}_c (t)+H.c.],\nonumber\\
\dot{\hat{P}}^{(2)}_c=&-\kappa \hat{P}^{(2)}_c+2g_m\mathcal{R}_e(E(t))[\cos(\omega_m t)\hat{X}_m+\sin(\omega_m t)\hat{P}_m]\nonumber\\
&-\sqrt{2}\kappa\mathcal{I}_m(E(t))-i\sqrt{\kappa}[e^{i\omega'_c t}\hat{\xi}_c (t)-H.c.],\nonumber\\
\dot{\hat{X}}_m=&-\gamma_m \hat{X}_m-2g_m\sin(\omega_m t)[\mathcal{R}_e(E(t))\hat{X}^{(2)}_c+\mathcal{I}_m(E(t))\hat{P}^{(2)}_c]\nonumber\\
&-g_m\sin(\omega_m t)|E(t)|^2+\sqrt{\gamma_m}[e^{i\omega_mt}\hat{\xi}_m(t)+H.c.],\nonumber\\
\dot{\hat{P}}_m=&-\gamma_m \hat{P}_m+2g_m\cos(\omega_m t)[\mathcal{R}_e(E(t))\hat{X}^{(2)}_c+\mathcal{I}_m(E(t))\hat{P}^{(2)}_c]\nonumber\\
&+g_m\cos(\omega_m t)|E(t)|^2-i\sqrt{\gamma_m}[e^{i\omega_mt}\hat{\xi}_m(t)-H.c.],
\label{eq:dm}
\end{align}
where $\mathcal{I}_m(E(t))$ ($\mathcal{R}_e(E(t))$) denotes the imaginary (real) part of the effect drive term in Eq. (\ref{def}), and $\hat{X}_m$ ($\hat{P}_m$) for the mechanical resonator is defined similarly as $\hat{X}_c$ ($\hat{P}_c$). 
As seen from Eq. (\ref{eq:dm}), the mechanical resonance frequency $\omega_m$ is relevant to this part of the cavity quadratures $\hat{X}^{(2)}_c$ and $\hat{P}^{(2)}_c$. There will be a detectable change in the cavity quadratures, if $\omega_m$ is changed by an added mass. 
In the latter discussions we only consider red-detuned drive field so that the system will stabilize for performing the measurements.

\section{mechanism for the mass sensing}
\label{sec:real}

In Fig. 2 we present the examples of how the change of mechanical frequency $\omega_m$ will cause the varied cavity quadratures $\langle\hat{X}_c\rangle$ and $\langle\hat{P}_c\rangle$, whose contributions from the process in Eq. (\ref{eq:dm}) are the only relevant ones to such change. As the mass of the mechanical resonator increases while the particles attach to it, its resonance frequency $\omega_m$ will be changed to $\omega_m-\Delta\omega_m$ according to the relation in Eq. (1). By the comparison of the measured quadratures for the changed and unchanged mechanical frequency, the mass of the attached particles can be deduced.
Without loss of generality, we let the device work at the room temperature $T=300K$, corresponding the thermal number $n_{th}=6\times10^4$ for $\omega_m=2\pi \times 100$MHz. The effects of the environmental temperature come from the initial state of the system (the state of the mechanical resonator in the initial thermal equilibrium with the environment) and some noise drive terms in Eq. (\ref{eq:dm}), but they do not contribute to the changes of the average quadratures $\langle\hat{X}_c(t)\rangle$ and $\langle\hat{P}_c(t)\rangle$, which are determined by the coherent drive terms in the dynamical equation. In principle the system can work at arbitrary temperature.

\begin{figure}[h!]
\centering
\includegraphics[width=\linewidth]{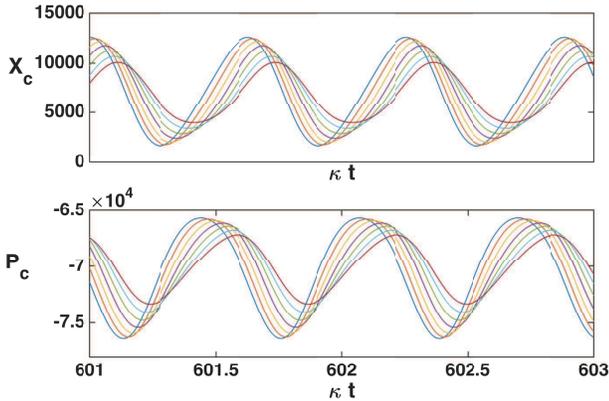}
\caption{Evolved cavity quadrature for the different mechanical frequency changes. The results are displayed in a reference frame in consistency with Eq. (\ref{eq:dm}). The largest amplitude is found for the original mechanical frequency with $\Delta \omega_m=0$. From this highest amplitude, the quadrature amplitude successively lowers with the increased mechanical frequency changes for $\Delta\omega_m/\kappa=0.001, 0.002,\cdots,0.006$. The system parameters are given as $g_m/\kappa=10^{-6}$, $\Delta/\kappa=100.01$, $\omega_m/\kappa=100$, $\gamma_m/\kappa=10^{-4}$, $n_{th}=6\times10^4$, and $E/\kappa=5\times10^6$.}
\label{fig2}
\end{figure}

The evolved quadratures $\langle\hat{X}_c(t)\rangle$ and $\langle\hat{P}_c(t)\rangle$ for the different deviations $\Delta \omega_m$ from 
the original mechanical frequency are illustrated in Fig. \ref{fig2}. The amplitude of the quadratures for the original mechanical frequency 
is found to be the largest. The cavity quadrature amplitudes can be detected with a homodyne-type detection; that is to mix the output field, which is proportional to the cavity field by the factor $\sqrt{\kappa}$, with a reference field of the same frequency, and their product is averaged by the integral with time so that the amplitude of $\hat{X}_c$ or $\hat{P}_c$ will be found by choosing the suitable phases of the reference field. In the examples of Fig. 2, the changed mechanical frequency can be detected up to the level of 
$\Delta\omega_m/\omega_m=10^{-5}$, corresponding to the mass change $\Delta m/m=10^{-5}$ via Eq. (1). The detection is realized by comparing the quadrature amplitudes after the system stabilizes. With a real-time homodyne detection \cite{rtq}, it is also possible to obtain 
the phase information of the cavity quadratures. Such method of detecting a tiny mass can work with a flexible drive field detuning $\Delta$, since in a realistic experiment the detuning of the drive field may not exactly match the original mechanical frequency $\omega_m$. In Fig. 2  the drive's detuning has a difference from the mechanical frequency by $\Delta-\omega_m=0.01\kappa$. 

\vspace{-0cm}

\section{sensor performance in the system's parameter space}
\label{sec:increase}

Next we investigate how the parameters of the system affect the sensor operation, so that one can choose the optimal ones. The changed cavity quadratures due to added mass are determined by the dynamical equations, Eq. (\ref{eq:dm}). We will find out the influence of the system parameters in the equations, which can be adjusted for the system, on the mass sensing. Since the two perpendicular quadratures' amplitudes are the same as seen from Fig. 2, we will only apply the quadrature $\hat{X}_c$ in the following discussions. We also consider a sufficiently long evolution time for the system so that the described quantities are stabilized ones. 

The performance of the mass sensing is measured by the quantity of the changed cavity quadrature 
$$\Delta X_c=|\langle \hat{X}_{c,M}(\omega_m+\Delta\omega_m)\rangle-\langle \hat{X}_{c,M}(\omega_m)\rangle|,$$ 
where $\langle X_{c,M}(\omega_m+\Delta\omega_m)\rangle$ and $\langle X_{c,M}(\omega_m)\rangle$ denote the peak values of the stabilized quadratures after and before more particles are attached to the mechanical resonator, respectively. 
Eq. (8) can be rewritten in a matrix form $\dot{\vec{x}}=\hat{M}\vec{x}+\vec{d}(t)+\vec{\hat{\xi}}(t)$, where $\vec{x}=(\hat{X}^{(2)}_c, \hat{P}^{(2)}_c, \hat{X}_m, \hat{P}_m)^T$, $\hat{M}$ is the dynamical matrix, $\vec{d}(t)$ represents the coherent drive terms proportional to $E/\Delta$, and $\vec{\hat{\xi}}$ represents the noise drives that are irrelevant to the evolved quadratures. In terms of this equation in the matrix form, one will have the changed quadrature vector as 
\begin{eqnarray}
\Delta \vec{x} &=&\int_0^t\big ({\cal T}e^{\int_\tau^t dt' \hat{M}(\omega_m+\Delta\omega_m,t')}-{\cal T}e^{\int_\tau^t dt' \hat{M}(\omega_m,t')}
\big)\vec{d}(\tau)d\tau\nonumber\\
&\approx &\int_0^t d\tau \int_\tau^t dt' \big(\hat{M}(\omega_m+\Delta\omega_m,t')-\hat{M}(\omega_m,t')\big)\vec{d}(\tau),\nonumber\\
\label{change}
\end{eqnarray}
where we expand the matrices $\hat{M}$ to the first order of the coefficient $J=g_mE/\Delta$, under the condition $J/\kappa\ll 1$ for the currently concerned situations. The relevant quadrature amplitude deviation $\Delta X_c$ for the sensor operation is obtained by eliminating the phase difference between $\hat{X}_{c}(\omega_m+\Delta\omega_m,t)$ and $\hat{X}_{c}(\omega_m,t)$, which give rise to $\Delta x_1$. Our numerical calculations will be based on the exact form on the first line of Eq. (\ref{change}). 

\subsection{The choices of the drive intensity and optomechanical coupling constant}

\begin{figure}[h!]
\centering
\includegraphics[width=\linewidth]{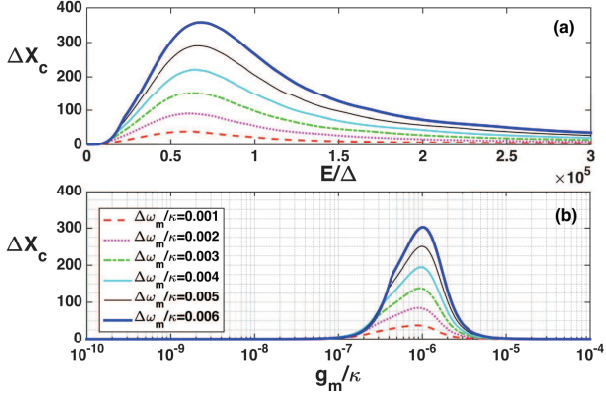}
\caption{(a) Amplitude changes of the cavity quadrature $\hat{X}_c$ as the functions the dimensionless drive intensity $E/\Delta$, for the different mechanical frequency change $\Delta\omega_m/\omega_m$ which are indicated by the legend in (b). The amplitudes are calculated in the range of the time $\kappa t=600-600.1$. Here we have a changed parameter $J=g_mE/\Delta$ along with the increased drive intensity $E$. (b): Quadrature amplitude changes with the single-photon coupling constant $g_m$, where we keep $E/\Delta$ fixed by using the parameters in Fig. \ref{fig2}.}
\label{fig4}
\end{figure}

The drive intensity $E$ is the parameter that can be conveniently adjusted for the system. We plot the quadrature variations with the drive intensity $E$ in Fig. 3(a), where the different curves show the $\Delta X_c$ for the different changes of the mechanical frequency $\omega_m$. There is an optimum value of $E$ to realize the largest quadrature difference $\Delta X_c$ for the sensor. Therefore, the best performance should be achieved by choosing the corresponding drive power. Certainly a more significant change $\Delta \omega_m$, which is proportional to the extra mass attached to the mechanical resonator, will induce a higher $\Delta X_c$ as shown in Fig. 3. Such optimal performance can be explained with the approximate variation in Eq. (\ref{change}). Since the parameter $J=g_m E/\Delta$ of the system is small (it is in the order of $10^{-2}\kappa$ in our currently concerned situations), its first order contribution becomes important 
so that the achieved variation $\Delta x_1$ in Eq. (\ref{change}) can be approximated by a polynomial function of $E$. An optimal value of $E$, which achieves the highest $\Delta X_c$, exists for such approximate form. In our numerical calculations we apply the exact form on the first line of Eq. (\ref{change}) to include the contributions from all orders of $J$, so the optimal drive intensities $E$ in Fig. 3 slightly differ from those of the corresponding polynomial functions of $E$.

The single-photon coupling constant $g_m$ indicating the interaction strength between the cavity mode and the mechanical mode is an important factor for the system, and usually it is challenging to realize a large value of this constant. We need to check what is a good value of $g_m$ so that the mass sensor operation can be well performed. In Fig. 3(b), we keep a fixed dimensionless drive intensity $E/\Delta$ to see how the coupling constant $g_m$ can affect the quadrature deviation $\Delta X_c$. Equivalently, that is to see how the parameter $J=g_m E/\Delta$ changes the system performance under a fixed effective drive intensity $E/\Delta$. In consistency with the previous discussions, we also find 
an optimal value of $g_m$ for the performance, which should be properly chosen in conjunction with other parameters.

Here are some examples for the choices of the parameters. If the mechanical frequency change is $\Delta\omega_m/\kappa=0.001$, the corresponding amplitude change will be about $\Delta X_c=40$, given $g_m/\kappa=10^{-6}$ and $E/\kappa=6\times10^6$. However, if one uses $g_m/\kappa=10^{-7}$ and $E/\kappa=6\times10^7$ (the same parameter $J$), the amplitude change will increase to $\Delta X_c=400$, which is better for the detection. A straightforward conclusion from Eq. (\ref{change}) is that, given a fixed parameter $J$, a higher drive intensity $E$ in $\vec{d}(t)$ will cause a more significant deviation $\Delta X_c$. 
Accordingly, in view of the results in Fig. 3, one can achieve an ultra-high sensor operation, e.g. $\Delta\omega_m/\omega_m=10^{-7}$ with $g_m/\kappa=10^{-8}$, or $\Delta\omega_m/\omega_m=10^{-8}$ with $g_m/\kappa=10^{-9}$, by increasing the drive intensity while lowering the constant $g_m$ so that the parameter $J$ is kept unchanged.

\subsection{The relations with the mechanical resonator's properties}

\begin{figure}[b!]
\centering
\includegraphics[width=\linewidth]{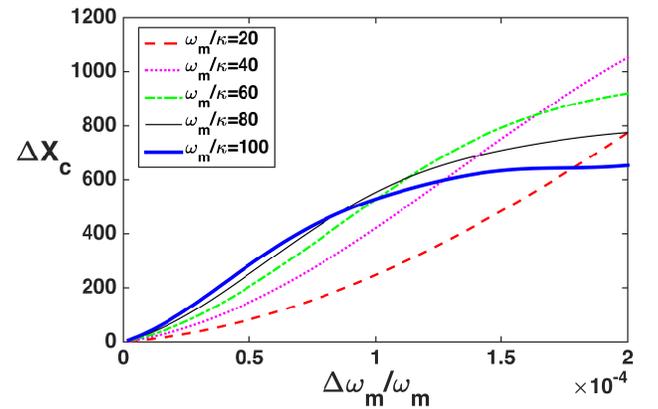}
\caption{Relations between the cavity quadrature change $\Delta X_c$ and the relative deviation 
$(\Delta \omega_m/\kappa)\times (\kappa/\omega_m)$ of the mechanical resonator's frequency. The drive field detuning matches the original mechanical frequency so that $\Delta=\omega_m$ for the different curves. We set $g_m=10^{-6}\kappa$, $\gamma_m=10^{-4}\kappa$, and $J=0.06$ to have a fixed $E/\Delta=J/g_m$ for all different curves.}
\label{}
\end{figure}

The mechanical resonator is an important component for the detector. One should know its relevant parameters for the design of the system. 
The first one is the sideband resolution $\omega_m/\kappa$ (the intrinsic frequency for the mechanical resonator) that is an essential parameter for many applications of optomechanically coupled systems. To show the effect of the resolved sideband parameter $\omega_m/\kappa$ in the concerned sensor operation, in Fig. 4 we illustrate how the cavity quadrature change $\Delta X_c$ responds to the relative deviation $\Delta\omega_m/\omega_m$ of the mechanical frequency, given the different values of $\omega_m/\kappa$. Here we suppose that the light field's detuning matches the mechanical resonator's original frequency, i.e. $\Delta=\omega_m$ before the particles attach to it. As shown in Fig. 4, the quadrature change $\Delta X_c$ after uploading the measured particles increases with $\omega_m/\kappa$ in the vicinity of $\Delta \omega_m/\omega_m=0$, tending to its limit value as $\omega_m/\kappa\rightarrow \infty$. Meanwhile we keep the fixed ratio $E/\Delta$ for the drive terms and coupling terms in Eq. (\ref{eq:dm}), so that the changed quadrature amplitude is purely due to the change of $\omega_m$. As the mechanical resonator's frequency lowers with the added masses $\Delta m$, the quadrature amplitude will immediately change with a quantity $\Delta X_c$. Such response becomes more sensitive for a higher value of $\omega_m/\kappa$ (the gradient of the curves at $\Delta\omega_m/\omega_m=0$ is larger for a higher $\omega_m/\kappa$ as shown in Fig. 4). So it is necessary 
to have high sideband resolution for the detection of a very small mass. Once there is more mass $\Delta m$ added, the lower value of $\omega_m/\kappa$ can even work better according to the results in Fig. 4.

\begin{figure}[t!]
\centering
\includegraphics[width=\linewidth]{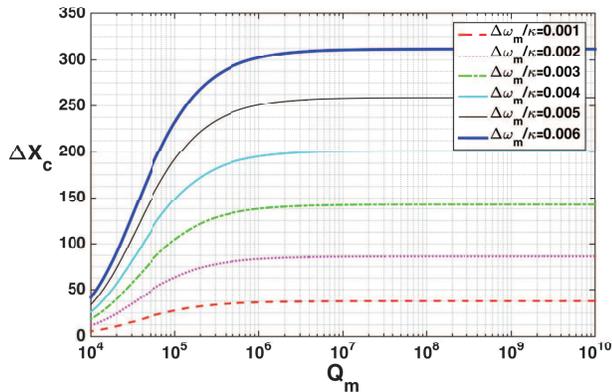}
\caption{Relations of the cavity quadrature $\hat{X}_c$ change with the quality factor of the mechanical resonator, for the different mechanical frequency deviations $\Delta\omega_m/\kappa$ that should be detected. The fixed parameters are the same as those in Fig. \ref{fig2}.}
\label{fig6}
\end{figure}

Finally, we check how good the mechanical quality factor should be in the operation. It is conceivable that a higher quality factor $Q_m=\omega_m/\gamma_m$, corresponding to a lower mechanical damping rate $\gamma_m$ [as a diagonal term in the matrix form of Eq. (\ref{eq:dm})] given a fixed $\omega_m$, will lead to the higher cavity and mechanical quadratures. However, the corresponding difference $\Delta X_c$ for the cavity quadrature due to a change $\Delta \omega_m$ in the mechanical frequency is not so straightforward. The mechanical damping rate $\gamma_m$ does not appear in the approximate form Eq. (\ref{change}) as in the first order of $J$, and it is from the higher-order corrections.
The actual relation between the $\Delta X_c$ and the mechanical quality factor is illustrated in Fig. \ref{fig6}. It shows that a higher quality factor for the mechanical resonator is better for the performance, but its improvement will saturate when the quality factor has been very large (the process has approached to the limit with no mechanical damping, i.e. $\gamma_m=0$). A quality factor in the order of $10^5-10^6$ can reach the best improvement for the examples in this paper. 

\section{Conclusion}
\label{sec:conclusion}
We have presented an approach to determining small
masses by means of detecting the change of cavity quadrature
for optomechanically coupled systems. There are the
following prominent advantages for the scheme: (1) the detector
operates at the room temperature; (2) the sensor operation
through detecting the light field's quadratures can be
sufficiently accurate, since the environmental noises do not
contribute to the evolved average quadratures (their effects are
averaged out in such detection); (3) the parameters for the systems
can be chosen flexibly, to be within those that have been
experimentally available. These features make the implementation
of the setup highly feasible. As we have shown with the detailed examples, the ultra-sensitive mass sensing, 
e.g., $\Delta m/m=10^{-8}-10^{-7}$, can be achieved by simply measuring the quadratures of light field. 
It is possible to detect a wide range of masses for nanoparticles with such a setup. 

\vspace{0.4cm}
\begin{acknowledgments}
The authors thank Prof. Yun-Feng Xiao for help discussions. This work is funded by National Natural Science Foundation of China 
(Grant No. 11574093, 61435007); Natural Science Foundation of Fujian Province of China (Grant No. 2017J01004); Promotion Program for Young and Middle-aged Teacher in Science and Technology Research of Huaqiao University (Grant No. ZQN-PY113). This research is also supported by the Arkansas High Performance Computing Center and the Arkansas Economic Development Commission.
\end{acknowledgments}

\end{document}